\documentclass[twocolumn,english,aps,prl,floatfix,showpacs]{revtex4}
\usepackage[T1]{fontenc}
\usepackage[latin1]{inputenc}

\makeatletter

\usepackage{babel}

\usepackage{babel}
\makeatother
\begin{document}

\title{Comment on {}``New Mean-Field Theory of the $tt^{\prime}t^{\prime\prime}J$
Model Applied to the High-T$_{c}$ Superconductors\char`\"{}}

\author{Alvaro Ferraz$^{1}$, Evgueny Kochetov$^{1,2}$, and Bruno Uchoa$^{1,3}$}

\affiliation{$^{1}$International Center of Condensed Matter
Physics, C.P. 04667, Brasilia, DF, Brazil}

\affiliation{$^{2}$Bogoliubov Theoretical Laboratory, Joint
Institute for Nuclear Research, 141980 Dubna, Russia}

\affiliation{$^{3}$Physics Department, Boston University, 590
Commonwealth Ave., Boston, MA 02215 }


\maketitle The authors of ref.$\,$\cite{wen} introduce a new
mean-field (MF) approach to the $t$-$J$ model of strongly
correlated electrons in terms of lattice spinons and doped
carriers, by constructing an enlarged Hilbert space with physical
and unphysical states. The purpose of this construction is clear:
to address the relaxation of the non-double occupancy (NDO)
constraint for the lattice electrons in a controlled way. In this
Comment, we show that the unphysical states are not properly
excluded and despite the claimed good agreement of their results
with experiment, the consistency of their MF theory is
unmotivated.

The original on-site 3d Hilbert space of the $t$-$J$ Hamiltonian
restricted by the NDO constraint includes a lattice electron
spin-up state, a lattice electron spin down-state, and a vacancy
state:
${\mathcal{H}}_{phys}:=\{|\uparrow\rangle,|\downarrow\rangle,|0\rangle\}$.
In order to recast such a model in terms of the spinon and
Gutzwiller projected dopon operators, the authors consider instead
the enlarged on-site Hilbert space,
${\mathcal{H}}^{enl}=\{|\uparrow,0\rangle,|\downarrow,0\rangle,|\uparrow,\uparrow\rangle,|
\uparrow,\downarrow\rangle,|\downarrow,\uparrow\rangle,|\downarrow,\downarrow\rangle\}$.
Here the first label denotes the up and down states of lattice
spins, whereas the second corresponds to the three dopon on-site
states, namely, the no dopon, spin-up and spin-down dopon states.
The physical space ${\mathcal{H}}_{phys}$ can then be identified
with the 3d subspace ${\mathcal{H}}_{phys}^{enl}$ by the following
covariant mapping,
$|\uparrow\rangle\rightarrow|\uparrow,0\rangle,\,|\downarrow\rangle\rightarrow|\downarrow,0\rangle,\,
|0\rangle\rightarrow\frac{1}{\sqrt{2}}(|\uparrow,\downarrow\rangle-|\downarrow,\uparrow\rangle).$
The unphysical states are
$\{|\uparrow,\uparrow\rangle,|\downarrow,\downarrow\rangle,\frac{1}{\sqrt{2}}(|\uparrow,\downarrow\rangle+|
\downarrow,\uparrow\rangle)\}={\mathcal{H}}_{unphys}^{enl}$. Under
this mapping, $|0\rangle\langle\uparrow|
\rightarrow\frac{1}{\sqrt{2}}(|\uparrow,\downarrow\rangle-|\downarrow,\uparrow\rangle)
\langle\uparrow,0|=\frac{1}{\sqrt{2}}\left[(\frac{1}{2}+S^{z}){\tilde{d}}_{\downarrow}^{\dagger}-S^{-}
{\tilde{d}}_{\uparrow}^{\dagger}\right]=:{\tilde{c}}_{\uparrow}.$
Although $\vec{S}$ and
$\tilde{d}_{\sigma}=d_{\sigma}(1-d_{-\sigma}^{\dagger}d_{-\sigma})$,
by definition, act in the whole of ${\mathcal{H}}^{enl}$, mixing
up physical and unphysical states, their specific combinations
given by the ${\tilde{c}}_{\sigma}$ operators act, as correctly
stated in~\cite{wen}, only on ${\mathcal{H}}_{phys}^{enl}$
annihilating all the unphysical states. This is because the
${\tilde{c}}_{\sigma}$ operators are by construction isomorphic to
the Hubbard operators $|0\rangle\langle\sigma|$, which satisfy the
NDO condition. However, the MF $t-J$ Hamiltonian cannot be written
in terms of ${\tilde{c}}_{\sigma}$ operators only and necessarily
acts in ${\mathcal{H}}^{enl}$. In this enlarged space the NDO
completeness relation given by
$\frac{1}{2}(|\uparrow,\downarrow\rangle-|\downarrow,\uparrow\rangle)(\langle\uparrow,\downarrow|-\langle\downarrow,
\uparrow|)+|\uparrow0\rangle\langle\uparrow0|+|\downarrow0\rangle\langle\downarrow0|\!=\!1$
just becomes a \textit{constraint}. It must be imposed to avoid
mixing of physical and unphysical states. In terms of the spinon
and dopon operators this constraint reads,
${\mathcal{C}}=\vec{J}^{2}-3/4(1-n)=0$. Here
$\vec{J}=\vec{S}+\vec{M}$ is the total spin on each lattice site,
and $\vec{M}={\tilde{d}}^{\dagger}\vec{\sigma}{\tilde{d}}$ and
$n={\tilde{d}}^{\dagger}{\tilde{d}}$ are the local dopon spin and
number operators, respectively.  This means that the on-site total
spin can either be $j=0$ or $j=1/2$, excluding the unphysical
states with $j=1$~\cite{hh}.

This constraint is missed in $\,$\cite{wen}. In contrast with the
exact representation given by their Eq.(2), the MF Hamiltonian (3)
mixes up both physical and unphysical sectors and the unphysical
states contribute to the MF phase diagram. Contrary to the
statement made in~\cite{rw}, the operator constraint
${\mathcal{C}}_{i}=0$ is not equivalent to the on-site conditions
$n_i\vec{J}_i=0$ in the enlarged space. As a consequence, in
contrast with the average constraint based on the completeness
relation, the conditions $<n_{i}\vec{J}_{i}>=0$ do not
discriminate between physical and unphysical states, making their
MF theory inconsistent~\cite{nJ}. Note finally that the unphysical
states may affect the dopon and spinon Green functions even beyond
the MF approximation. For example, for any
$H=H({\tilde{c}}_{\sigma}^{\dagger},{\tilde{c}}_{\sigma}$) one
gets,
$tr({\tilde{d}}_{\uparrow}^{\dagger}(t){\tilde{d}}_{\uparrow})
=tr(e^{-itH}{\tilde{d}}_{\uparrow}^{\dagger}e^{itH}{\tilde{d}}_{\uparrow})
=tr({\tilde{d}}_{\uparrow}^{\dagger}(t){\tilde{d}}_{\uparrow})_{{\mathcal{C}}=0}
+\langle\uparrow\uparrow|(e^{-itH}{\tilde{d}}_{\uparrow}^{\dagger}e^{itH}{\tilde{d}}_{\uparrow})
|\uparrow\uparrow\rangle
=tr({\tilde{d}}_{\uparrow}^{\dagger}(t){\tilde{d}}_{\uparrow})_{{\mathcal{C}}=0}+\langle\uparrow,0|e^{-itH}
|\uparrow,0\rangle,$ demonstrating that the unphysical state
$|\uparrow,\uparrow\rangle$ produces a non-trivial contribution if
not excluded by the constraint.


\begin{thebibliography}{1}
\bibitem{hh} In the undoped case $(n_i=0)$ this constraint reduces to
the equation $\vec S^2_i=3/4$ which is appropriate for the
conventional $s=1/2$ Heisenberg model.
\end{thebibliography}
\end{document}